\documentclass[aps,superscriptaddress, showpacs,preprintnumbers,  nofootinbibt,twocolumn]{revtex4}
\usepackage{amssymb}
\usepackage{eurosym}
\usepackage{graphicx}
\usepackage{amsfonts}
\usepackage{amssymb}
\usepackage{amsmath}
\usepackage{bm}
\usepackage{color}

\def\be{\begin{equation}}
\def\ee{\end{equation}}
\def\bea{\begin{eqnarray}}
\def\eea{\end{eqnarray}}

\begin{document}

\title{Exact Scalar-Tensor Cosmological Models}
\author{J. A. Belinch\'{o}n}
\email{jose.belinchon@uda.cl}
\affiliation{Dept. F\'{\i}sica, Facultad Ciencias Naturales, Universidad de Atacama,
Copayapu 485, Copiap\'o, Chile}
\author{T. Harko}
\email{t.harko@ucl.ac.uk}
\affiliation{Department of Physics, Babes-Bolyai University, Kogalniceanu Street,
Cluj-Napoca 400084, Romania,}
\affiliation{Department of Mathematics, University College London, Gower Street, London
WC1E 6BT, United Kingdom}
\author{M. K. Mak}
\email{mankwongmak@gmail.com}
\affiliation{School of Science and Technology, The Open University of Hong Kong, \\ Homantin, Kowloon, Hong Kong, P. R. China}
\affiliation{Departamento de F\'{\i}sica, Facultad de Ciencias Naturales, Universidad de
Atacama, Copayapu 485, Copiap\'o, Chile}
\date{\today }

\begin{abstract}
Scalar-tensor gravitational theories are important extensions of standard
general relativity, which can explain both the initial inflationary
evolution, as well as the late accelerating expansion of the Universe. In
the present paper we investigate the cosmological solution of a scalar-tensor
gravitational theory, in which the scalar field $\phi $ couples to the
geometry via an arbitrary function $F(\phi $). The kinetic energy of the
scalar field as well as its self-interaction potential $V(\phi )$ are also
included in the gravitational action. By using a standard mathematical
procedure, the Lie group approach, and Noether symmetry techniques, we
obtain several exact solutions of the gravitational field equations
describing the time evolutions of a flat Friedman-Robertson-Walker Universe
in the framework of the scalar-tensor gravity. The obtained solutions can
describe both accelerating and decelerating phases during the cosmological
expansion of the Universe.
\end{abstract}

\maketitle



\section{ Introduction}

There is strong observational and theoretical evidence that general
relativity (GR) is not the complete, and final geometric theory describing
the nature of the gravitational force. From the theoretical point of view,
there are many shortcomings of GR in cosmology and quantum field theory. For
example, it is known that the standard model of cosmology based on GR and on
particle physics is unable to depict the gravitational force in extreme
regimes, including the description of gravity on microscopic scales of the
order of Planck length. In order to describe the very early expansion
history of the universe accurately, one needs to quantize the geometry, in a
way to connect it with quantum field theory. On the other hand, from the
particle physics point of view, we must note the fact that the Dirac
equation is extremely successful in unifying special relativity and quantum
mechanics. Although the unification of GR and quantum field theory has
gained some progresses, yet the unification of GR and of quantum mechanics
still remains an open question.

On the other hand, observational evidence obtained on distant Type Ia
Supernovae (SNeIa) between 1998 to 2010, suggested that the universe has
undergone a late time accelerated expansion \cite{1,2,3,4}. Recently, the
Cosmic Microwave Background temperature anisotropies measured by the
Wilkinson Microwave Anisotropy Probe and the Planck satellites have been
reported in \cite{5,6}, confirming that the total amount of baryonic matter
in the Universe is very low. It is worth to notice that the study of the
baryonic acoustic oscillations have also confirmed, and correlated the other
cosmological observations \cite{7}.

In order to explain the accelerated expansion of the universe, one may
introduce the idea of dark energy (DE), which from a mathematical point of
view and in the context of GR represents a new contribution in the matter
part of the Einstein's gravitational field equation. The most popular
candidate for dark energy is the cosmological constant $\Lambda $, which is
introduced without modifying the geometric part of the Einstein field
equations. On the other hand, dark energy can also be explained as a
modification of the geometrical part of Einstein's equation. Such an
approach leads to the so called modified gravitational theories (see reviews
in \cite{8,9,10,11,12,13,14}).

Among the most interesting modified gravity theories are the scalar-tensor
theories of gravitation, which are an important generalization of standard
GR theory. In the so-called Brans-Dicke theory, an additional scalar field $%
\phi $ besides the metric tensor $g_{\mu \nu }$ and a dimensionless coupling
constant $\omega $ were introduced in order to describe the gravitational
interaction \cite{BD,B1,B2,B3,B4}. Brans-Dicke theory recovers the results
of GR for large value of the coupling constant $\omega $, that is for $%
\omega >500$. Similar formalisms were developed earlier by Jordan \cite%
{J1,J2}. Note that the history and developments of scalar-tensor theories
(STT) can be found in \cite{h1,h2,h3,h4}.

The astrophysical and cosmological implications of the Brans-Dicke type
theories have been extensively investigated. The dynamics of a causal bulk
viscous cosmological fluid filled flat homogeneous Universe in the framework
of the BD theory was reported in \cite{HM}. Scalar-tensor theories are very
promising because they can provide an explanation of the inflationary
behavior in cosmology. Also the scalar field can be considered an additional
degree of freedom of the gravitational interaction. Scalar-tensor theories
has been used to model DE, because the scalar fields are a good candidate
for phantom and quintessence fields \cite{peeble,TLM,TM1}. Furthermore, it
is natural to couple the scalar field to the curvature after
compactification of higher dimensional theories of gravity such as
Kaluza-Klein and string theory, offering the possibility of linking
fundamental scalar fields with the nature of the DE.

Very recently, Hojman symmetry approach was used to study the behavior of
the flat Friedman-Robertson-Walker (FRW) universe in the context of STT \cite%
{1a}. The possibility of probing a yet unconstrained region of the parameter
space of STTs is based on the fact that stability properties of highly
compact neutron stars may radically differ from those in GR \cite{R}. In the
scalar-tensor model with Gauss-Bonnet and kinetic couplings the power law
dark energy solution may be described by Higgs-type potential \cite{LN}. A
covariant multipolar Mathisson-Papaetrou-Dixon type approach was used to
derive the equations of motion in a systematic way for both Jordan and
Einstein formulations of STT in \cite{Y}. The parametrized post-Newtonian
(PPN) parameters $\gamma $ and $\beta $ for general STT was computed, and it
was suggested that the PPN parameters $\gamma $ and $\beta $ given by the
condition $\left\vert \gamma -1\right\vert \sim $ $\left\vert \beta
-1\right\vert \sim 10^{-6}$ may be detectable by a satellite that carries a
clock with fractional frequency uncertainty $\frac{\Delta f}{f}\sim 10^{-16}$
in an eccentric orbit around the Earth \cite{PPN}.

A complete Noether symmetry analysis in the framework of scalar-tensor
cosmology was reported in \cite{NS} (see also \cite{MNSa,MNSb} for other
approaches). Scalar-tensor cosmology with $1/R$ curvature correction by
Noether symmetry was discussed in \cite{NN}. The weak field limit of STT of
gravitation was discussed in view of conformal transformations in \cite{A},
and a new reconstruction method of STT based on the use of conformal
transformations was proposed. This method allows the derivation of a set of
interesting exact cosmological solutions in BD gravity, as well as in other
extensions of GR \cite{S}. Recently, (An)isotropic models in the presence of
variable gravitational and cosmological constants were studied in the
context of scalar-tensor cosmology \cite{Tony}. Generalized self-similar STT
was investigated and some new exact self-similar solutions were obtained in
\cite{Tony1}, by using the Kantowski-Sach models. Exact solution of
gravitational field equations describing the dynamics of the anisotropic
universe with string fluid was presented in the framework of STT in \cite{AK}%
. Testing the feasibility of scalar-tensor gravity by scale dependent mass
and coupling to matter was investigated in \cite{DF}. The phase space of
Friedmann-Lemaitre-Robertson-Walker models derived from STT in the presence
of the non-minimal coupling $F\left( \phi \right) =\xi \phi ^{2}$ and the
effective potential $V\left( \phi \right) =\lambda \phi ^{n}$ was studied\
in \cite{FS}. The Noether symmetries of a generalized scalar-tensor, Brans-Dicke type cosmological model, in which the explicit scalar field dependent couplings to the Ricci scalar, and to the scalar field kinetic energy, respectively, were considered, was investigated in \cite{BHM}. The scalar field self-interaction potential into the gravitational action was also included. Three cosmological solutions describing the time evolution of a spatially flat Friedman-Robertson-Walker Universe were obtained, and the cosmological properties of the solutions were investigated in detail. The obtained models can describe a large variety of cosmological evolutions, including models that experience a smooth transition from a decelerating to an accelerating phase.

The purpose of the present paper is to consider a detailed mathematical
analysis of the cosmological solutions of a special class of scalar-tensor
type gravitational theories, in which in the gravitational action the Ricci
scalar couples to an arbitrary function of the scalar field $\phi$. The
scalar field kinetic energy, as well as the scalar field self interaction
potential are also considered. As a first step in our analysis we present
several exact analytical solutions of the gravitational field equations,
describing the time evolutions of the flat FRW universe in the presence of
the scalar field by using the standard procedure. Then two rigorous
mathematical approaches, the Lie group approach, and the Noether symmetry
techniques, respectively, are applied for the investigation of the
gravitational field equations. Both approaches provide several classes of
solutions of the gravitational field equations, allowing the determination
of the cosmological scalar field and of its self-interaction potential.

The present paper is organized as follows. The scalar-tensor gravitational
model and the cosmological field equations are presented in Section~\ref%
{sect2}. A set of simple analytical solutions of the gravitational field
equations describing the time evolution of the flat FRW universe in the
presence of the scalar field by using standard procedures are obtained in
Section~\ref{sect3}. The cosmological field equations are investigated by
using the Lie group approach in Section~\ref{sect4}. The Noether symmetry
techniques are applied in Section~\ref{sect5}. We conclude our results in
Section~\ref{sect6}.

\section{Scalar-Tensor Cosmology}

\label{sect2}

In the present paper we are considering a gravitational model in which the
Ricci scalar, describing pure gravity, couples in a non-standard way with a
scalar field. For this model the gravitational action takes the form \cite%
{1a,1c}
\begin{equation}
S=\int d^{4}x\sqrt{-g}\left[ F\left( \phi \right) R+\frac{1}{2}g^{\mu \nu
}\nabla _{\mu }\phi \nabla _{\nu }\phi -V\left( \phi \right) \right] ,
\label{1}
\end{equation}%
where $R$ is the Ricci scalar. The plus sign appears in the kinetic term of
the action (\ref{1}) corresponding to the phantom phase \cite{p1,p2}. The
arbitrary functions $F\left( \phi \right) $ and $V\left( \phi \right) $ are
the coupling to the Ricci scalar and the potential of the scalar field $\phi
$ respectively. In a spatially flat FRW space-time, with metric
\begin{equation}
ds^{2}=dt^{2}-a^{2}(t)\left( dx^{2}+dy^{2}+dz^{2}\right) ,
\end{equation}%
the Lagrangian density on the configuration space $\left( a,\phi \right) $
for non-minimal coupled scalar-tensor cosmology is%
\begin{equation}
\mathcal{L}=6F\left( \phi \right) a\dot{a}^{2}+6F^{^{\prime }}\left( \phi
\right) a^{2}\dot{a}\dot{\phi}+\frac{1}{2}a^{3}\dot{\phi}^{2}-a^{3}V\left(
\phi \right) ,  \label{2}
\end{equation}%
where the overdot and the prime denote the derivative with respect to the
time and the scalar field $\phi $, respectively. The function $a(t)$ is the
scale factor of the Universe, which gives the full information on the
expansion history of the Universe. Now, the Euler-Lagrangian equations can
be obtained from Eq. (\ref{2}), and are given by \cite{1a}%
\begin{equation}
2\dot{H}+3H^{2}+\left( 2H\dot{\phi}+\ddot{\phi}\right) \frac{F^{^{\prime }}}{%
F}+\left( F^{^{\prime \prime }}-\frac{1}{4}\right) \frac{\dot{\phi}^{2}}{F}=-%
\frac{V}{2F},  \label{3}
\end{equation}%
\begin{equation}
\ddot{\phi}+3H\dot{\phi}+6\left( \dot{H}+2H^{2}\right) F^{^{\prime
}}+V^{^{\prime }}=0,  \label{4}
\end{equation}%
corresponding to the Einstein-Friedmann equation, and to the Klein-Gordon
equation for the FRW geometry, respectively. In Eqs.~(\ref{3}) and (\ref{4})
we have defined the Hubble parameter $H$ as $H=\frac{\dot{a}}{a}$. The
energy function corresponding to the Einstein $\left( 0,0\right) $ equation
takes immediately the form
\begin{equation}
-6H\left( FH+F^{^{\prime }}\dot{\phi}\right) -\frac{\dot{\phi}^{2}}{2}=V.
\label{5}
\end{equation}%
By eliminating the potential $V$ from Eqs. (\ref{3}) and (\ref{5}), we get
the general differential equation given by
\begin{equation}
H^{^{\prime }}=\frac{F^{^{\prime }}}{2F}H+\frac{\dot{\phi}}{4F}\left(
1-2F^{^{\prime \prime }}\right) -\frac{\ddot{\phi}}{2\dot{\phi}}\frac{%
F^{^{\prime }}}{F}.  \label{6}
\end{equation}%
By defining the arbitrary function $M\left( \phi \right) $ as $M\left( \phi
\right) =\dot{\phi}$, with this transformation, thus Eq. (\ref{6}) becomes%
\begin{equation}
H^{^{\prime }}=\frac{F^{^{\prime }}}{2F}H+\frac{M}{4F}\left( 1-2F^{^{\prime
\prime }}\right) -\frac{F^{^{\prime }}}{2F}M^{^{\prime }}.  \label{7}
\end{equation}%
In the following, we shall present the analytical solutions of Eq.~(\ref{7}%
), which describe the dynamics of the flat FRW Universe in the presence of
the scalar field coupled to the geometrical part of the action.

\section{Analytical solutions of the gravitational field equations}

\label{sect3}

In order to obtain a solution of the differential Eq.~(\ref{7}), we assume
that the coupling function $F\left( \phi \right) $, the function $M\left(
\phi \right) $ and the Hubble function $H\left( \phi \right) $ obey the
relations

\begin{equation}
F\left( \phi \right) =\alpha _{0}\phi ^{2},  \label{a1}
\end{equation}%
\begin{equation}
M\left( \phi \right) =\alpha _{1}\phi ^{s},  \label{a2}
\end{equation}%
\begin{equation}
H\left( \phi \right) =\alpha _{2}\phi ^{s-1},  \label{a3}
\end{equation}%
respectively, where $s$ and $\alpha _{i}$, $i=0,1,2$ are arbitrary
constants. By substituting Eqs. (\ref{a1},\ref{a2},\ref{a3}) into Eq. (\ref%
{7}), the latter yields the relation
\begin{equation}
\alpha _{2}\left( s-2\right) =\alpha _{1}\left( \frac{1}{4\alpha _{0}}%
-1-s\right) .  \label{a4}
\end{equation}%
By substituting Eqs. (\ref{a1}), (\ref{a2}, and (\ref{a3}) into Eq.~ (\ref{5}%
), the latter gives the potential as
\begin{eqnarray}
V\left( \phi \right) &=&-\left( 6\alpha _{0}\alpha _{2}^{2}+12\alpha
_{0}\alpha _{1}\alpha _{2}+\frac{1}{2}\alpha _{1}^{2}\right) \phi ^{2s}
\notag \\
&=&\frac{\alpha _{1}^{2}\left( 12\alpha _{0}-1\right) \left[ 3+4\alpha
_{0}\left( s+1\right) \left( s-5\right) \right] }{8\alpha _{0}\left(
s-2\right) ^{2}}\phi ^{2s},
\end{eqnarray}%
or, equivalently,
\begin{widetext}
\bea
\frac{V(\phi)}{2\alpha _{0}\alpha _{2}^{2}}&=&\left\{ \frac{4\alpha _{0}\left[ 11+\left(
2-s\right) s+12\alpha _{0}\left( s-3\right) \left( 1+s\right) \right] -3}{%
\left[ 1-4\alpha _{0}\left( 1+s\right) \right] ^{2}}\right\}
\phi ^{2s},
\label{a5}
\end{eqnarray}%
\end{widetext}where we have used Eq. (\ref{a4}) in Eq. (\ref{5}). In order
not to have a vanishing potential $V\left( \phi \right) $, the coupling
constant $\alpha _{0}$ must satisfy in this cosmological model the following
relations
\begin{equation}
\alpha _{0}\neq \frac{1}{12},\alpha _{0}\neq \frac{3}{4\left( s+1\right)
\left( 5-s\right) },  \label{b1}
\end{equation}%
\begin{equation}
\alpha _{0\pm }\neq \frac{s^{2}-2s-11\pm \left( s-1\right) \sqrt{s^{2}-2s+13}%
}{24\left( s-3\right) \left( s+1\right) }.  \label{b3}
\end{equation}%
Now with the help of $M\left( \phi \right) =\dot{\phi}$, then Eq. (\ref{a2})
can be integrated to give
\begin{equation}
\phi \left( t\right) =\phi _{1}\left( t-t_{0}\right) ^{\frac{1}{1-s}},s\neq
1,  \label{a6}
\end{equation}%
where $t_{0}$ is an arbitrary constant of integration, and we have denoted
the arbitrary constant $\phi _{1}$ as $\phi _{1}=\left[ \alpha _{1}\left(
1-s\right) \right] ^{\frac{1}{1-s}}$. By inserting Eq. (\ref{a6}) into Eq. (%
\ref{a3}), thus the latter can be integrated to yield the scale factor as
\begin{equation}
a\left( t\right) =a_{0}\left( t-t_{0}\right) ^{\mu },s\neq 1,2,  \label{a7}
\end{equation}%
where $a_{0}$ is an arbitrary constant of integration, and we have denoted
the arbitrary constant $\mu $ as
\begin{equation}
\mu =\frac{\alpha _{2}}{\alpha _{1}\left( 1-s\right) }=\frac{1+s-\frac{1}{%
4\alpha _{0}}}{\left( s-2\right) \left( s-1\right) }.
\end{equation}

If the constant $\alpha _{0}$ satisfies the relation $\alpha _{0}=\frac{1}{%
4\left( s+1\right) }$, then the FRW line element will reduce to the static
case. Therefore we have completely obtained a cosmological solutions of the
STT model, given by Eqs. (\ref{a5}), (\ref{a6}), (\ref{a7}), and describing
the time evolution of the flat FRW universe in the presence of the scalar
field $\phi $. In the next Section, we shall present the analytical
solutions of the differential Eq. (\ref{6}) by means of the Lie approach,
subsequently leading to the complete solutions of the gravitational field
equation describing the dynamics of the FRW universe in the presence of the
scalar field in the framework of the STT.

\section{Standard Lie Group Approach}

\label{sect4}

It is known that a vector field $X$ defined as
\begin{equation}
X=\xi (x,y)\frac{\partial }{\partial x}+\eta (x,y)\frac{\partial }{\partial y%
},
\end{equation}%
where $\xi (x,y)$ and $\eta (x,y)$ are two arbitrary functions is a symmetry
of a differential equation, which is an invertible transformation that
leaves it form-invariant. By applying the standard Lie procedure, one needs
to solve the following over-determined system of the partial differential
equations for the functions $\eta $ and $\xi $, based on the extended
infinitesimal or prolonged transformations. Thus this approach allows us to
determine the set of the symmetries admitted by the considered equation. By
considering a general second order differential equation,
\begin{equation}
\phi ^{\prime \prime }=\psi (t,\phi ,\phi ^{\prime }),\text{\ }
\label{lieexam}
\end{equation}%
then we apply the standard procedures of Lie group analysis to Eq. (\ref%
{lieexam}) \cite{Bluman}. A vector field $X$ defined as
\begin{equation}
X=\xi (t,\phi )\partial _{t}+\eta (t,\phi )\partial _{\phi },
\end{equation}%
is a symmetry of Eq. (\ref{lieexam}) if the arbitrary functions $\xi (x,y)$
and $\eta (x,y)$ satisfy the partial differential equation
\begin{equation*}
-\xi f_{t}-\eta f_{\phi }+\eta _{tt}+\left( 2\eta _{t\phi }-\xi _{tt}\right)
\phi ^{\prime }+\left( \eta _{\phi \phi }-2\xi _{t\phi }\right) \phi
^{\prime 2}-\xi _{\phi \phi }\phi ^{\prime 3}
\end{equation*}%
\begin{equation}
+\left( \eta _{\phi }-2\xi _{t}-3\phi ^{\prime }\xi _{\phi }\right) f-\left[
\eta _{t}+\left( \eta _{\phi }-\xi _{t}\right) \phi ^{\prime }-\phi ^{\prime
2}\xi _{\phi }\right] f_{\phi ^{\prime }}=0.  \label{ber2}
\end{equation}%
The knowledge of one symmetry $X$ provides the form of a particular solution
as an invariant of the operator $X$, i.e. a solution of the differential
equation $\frac{dx}{\xi (x,y)}=\frac{dy}{\eta (x,y)}$. This particular
solution is known as an invariant solution (generalization of similarity
solution). As a first step in our analysis of STT cosmologies, we rewrite
Eq. (\ref{6}) in the form%
\begin{equation}
\ddot{\phi}=\frac{1}{F^{^{\prime }}}\left[ \left( \frac{1}{2}-F^{^{\prime
\prime }}\right) \dot{\phi}^{2}-2F\dot{H}+F^{\prime }\dot{\phi}H\right] .
\label{eq2}
\end{equation}%
By applying the Lie group approach to Eq. (\ref{eq2}), we immediately obtain
the system of the partial differential equations%
\begin{widetext}
\begin{align}
-2F^{\prime 2}\xi ^{\prime \prime }+2F^{\prime }\left( 2F^{\prime \prime
}-1\right) \xi ^{\prime }& =0,  \label{sys1} \\
2F^{\prime 2}\left( \eta ^{\prime \prime }-2\dot{\xi}^{\prime }+2H\xi
^{\prime }\right) +\left( 2F^{\prime }F^{\prime \prime \prime }-2F^{\prime
\prime 2}+F^{\prime \prime }\right) \eta +F^{\prime }\left( 2F^{\prime
\prime }-1\right) \eta ^{\prime }& =0,  \label{sys2} \\
12FF^{\prime }\dot{H}\xi ^{\prime }+2F^{\prime 2}\left( 2\dot{\eta}^{\prime
}-\ddot{\xi}-\left( H\dot{\xi}+\dot{H}\xi \right) \right) +2F^{\prime
}\left( 2F^{\prime \prime }-1\right) \dot{\eta}& =0,  \label{sys3} \\
4\left( F^{\prime 2}-FF^{\prime \prime }\right) \dot{H}\eta +4FF^{\prime
}\left( 2\dot{\xi}\dot{H}+{\ddot{H}}\xi -4\eta ^{\prime }\right) +2F^{\prime
}\left( \ddot{\eta}-\dot{\eta}H\right) & =0.  \label{sys4}
\end{align}
\end{widetext}

In the following, we shall find two analytical solutions of Eq. (\ref{eq2})
depicting the dynamics of the FRW universe, with the help of the system
given by Eq. (\ref{sys1})-(\ref{sys4}), subsequently leading to some classes
of solutions of the gravitational field equations.

\subsection{Solution 1}

Firstly, by imposing the symmetry $\left[ ct,\phi \right] $ where $c\in
\mathbb{R},$ and using Eqs. (\ref{sys1})-(\ref{sys4}), we obtain the
coupling function $F\left( \phi \right) $, the Hubble parameter $H\left(
t\right) $ and the scale factor $a(t)$ as
\begin{equation}
F=m\phi ^{2},H=\frac{a_{1}}{t},a(t)=t^{a_{1}},  \label{kk}
\end{equation}%
respectively where $m$,$a_{1}\in \mathbb{R}$. Now by inserting these results
into Eq.(\ref{eq2}), we obtain the scalar field $\phi \left( t\right) $ in
the form
\begin{equation}
\phi \left( t\right) =\phi _{0}t^{S}\left( c_{1}t^{B}-c_{2}\right) ^{\alpha
},  \label{pp}
\end{equation}%
where we have defined the arbitrary constants $S$,$B$,$\phi _{0}$ and $%
\alpha $ as%
\begin{eqnarray}
B &=&\sqrt{a_{1}^{2}+\left( 10-\frac{1}{m}\right) a_{1}+1},2S=\alpha \left(
1+a_{1}-B\right) ,  \notag \\
&&\phi _{0}=\left( B\alpha \right) ^{\pm \alpha },\alpha =\frac{4m}{8m-1},
\end{eqnarray}%
respectively, with $c_{1},c_{2}\in \mathbb{R}$. Note that the invariant
solution induced by the symmetry $\left[ t,n\phi \right] $ yields the
relation
\begin{equation}
\phi =\phi _{0}t^{n},  \label{ISLG1}
\end{equation}%
where we have defined the arbitrary constant $n$ as $2n=\alpha \left(
1+a_{1}\pm B\right) $. One may find the exact value of constant $n$ from Eq.
(\ref{eq2}). For simplicity, we assume that $c_{1}=0$, then from the field
Eq. (\ref{5}), we obtain the potential $V$ as
\begin{equation}
V\sim t^{\gamma },  \label{mk}
\end{equation}%
the proportionality constant obtained from Eq. (\ref{mk}) is complicated
thus we shall not present its explicit form here, where we have denoted the
arbitrary constant $\gamma $ as
\begin{equation}
\gamma =\frac{2}{8m-1}\left[ 2m\left( a_{1}-3-B\right) +1\right] .
\end{equation}%
Note that the invariant solution is homothetic, since it verifies the
relation obtained in the matter collineation approach, that is,
\begin{equation}
L_{\mathbb{H}}\left( F^{-1\left( \phi \right) }T_{\mu \nu }\right) =0,
\end{equation}%
where $\mathbb{H}$ is the homothetic vector field associated to the flat FRW
metric, $L$ is the Lie derivative, and the energy momentum tensor $^{\left(
\phi \right) }T_{\mu \nu }$ is defined as
\begin{equation}
^{\left( \phi \right) }T_{\mu \nu }=\phi _{;\mu }\phi _{;\nu }-\frac{1}{2}%
g_{\mu \nu }\phi _{;i}\phi ^{;i}+g_{\mu \nu }V.
\end{equation}%
In view of the above equations we get the relations: $F^{-1}\phi ^{\prime
2}=t^{-2}$ and $F^{-1}V=t^{-2}$, which are verified for our invariant
solution.

\begin{figure}[h!]
\begin{center}
\includegraphics[height=2in,width=3in]{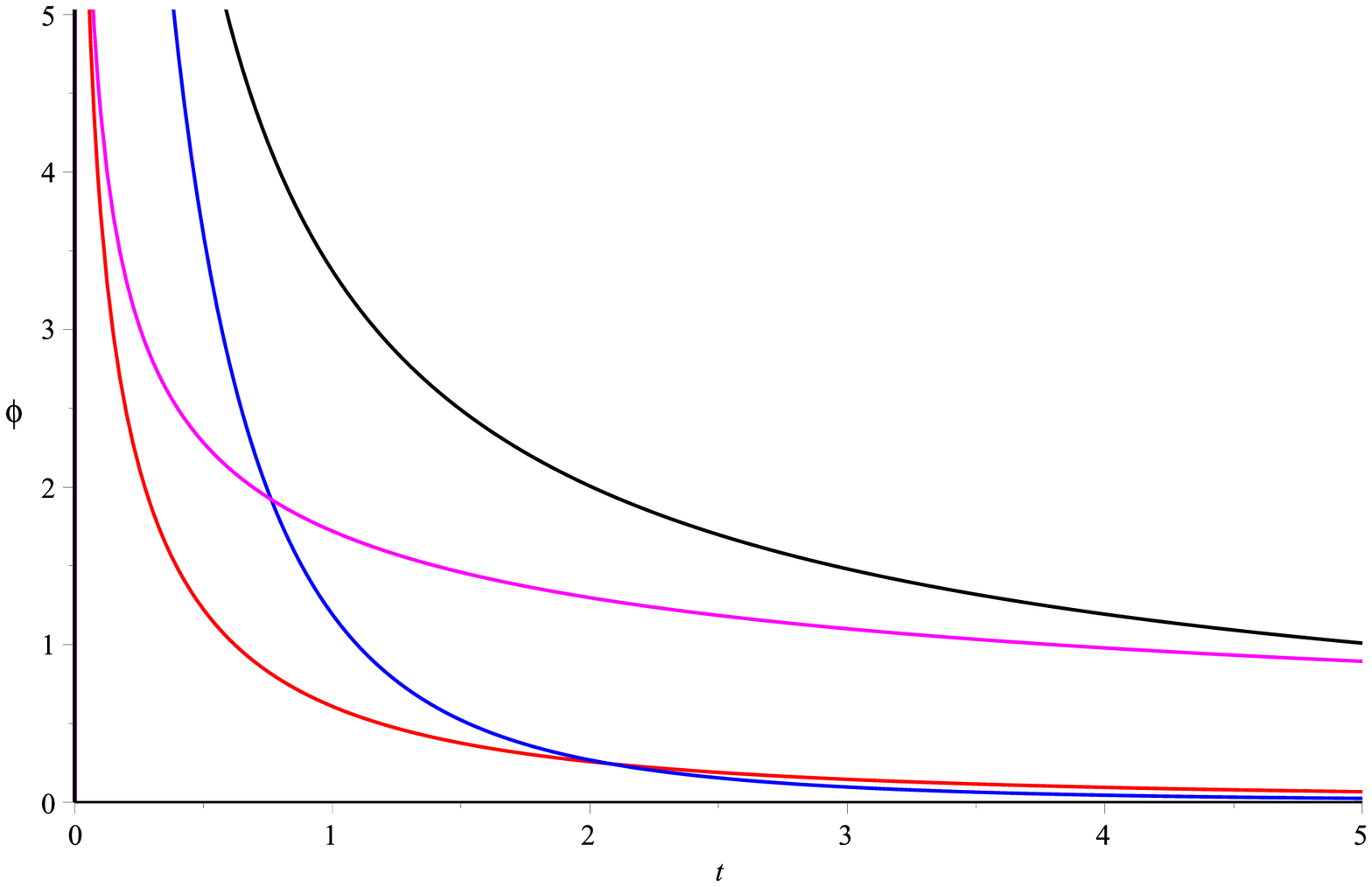} %
\includegraphics[height=2in,width=3in]{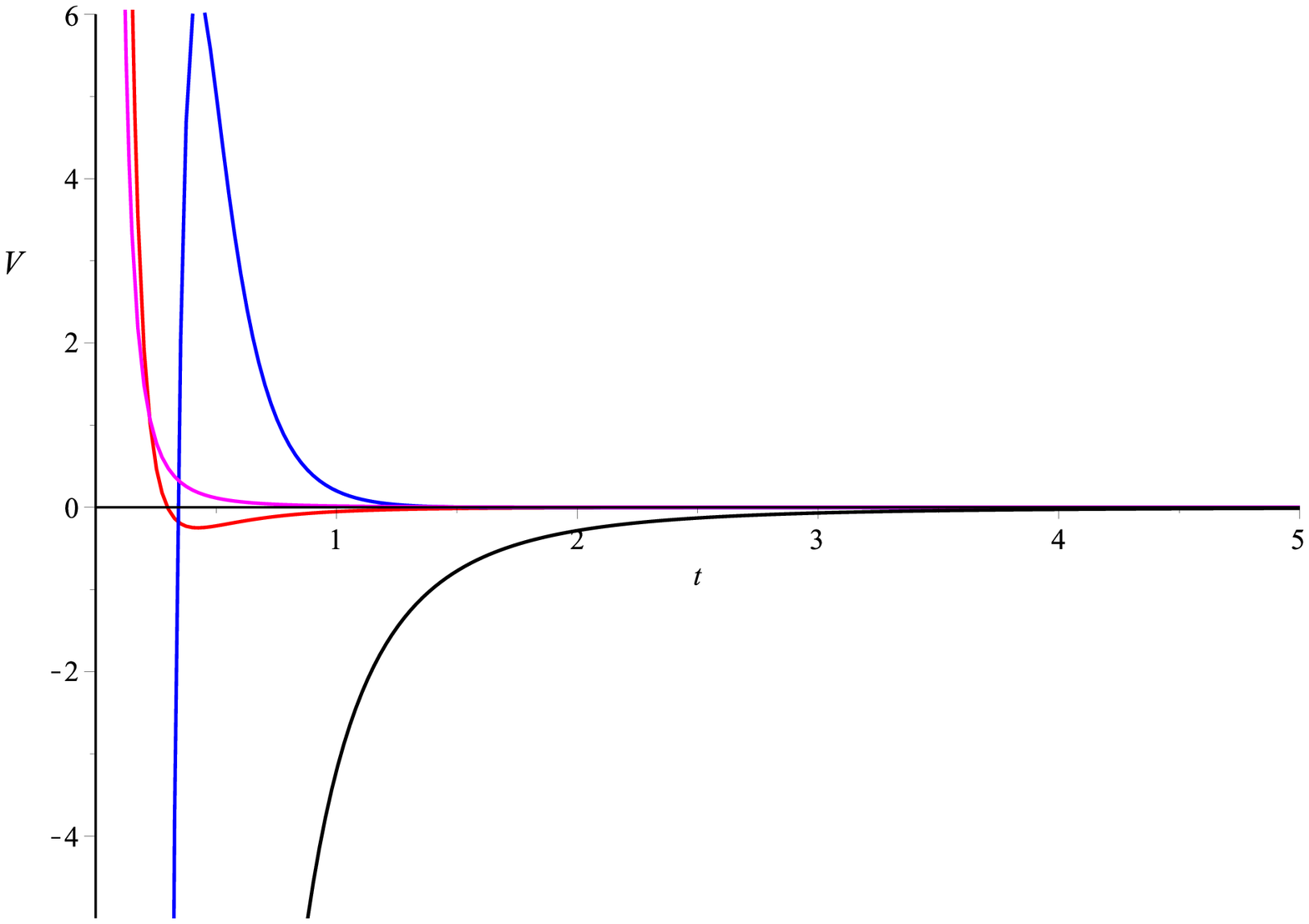}
\end{center}
\caption{Plots of $\protect\phi$ and $V$. Red line: $%
c_{1}=1,c_{2}=-1,m=3/32,a_{1}=1/2$. Blue line: $%
c_{1}=1,c_{2}=-1,m=3/32,a_{1}=3/2$. Magenta line: $%
c_{1}=0,c_{2}=1,m=3/32,a_{1}=1/2$ and black line: $%
c_{1}=0,c_{2}=1,m=3/32,a_{1}=3/2$. }
\label{Fig1}
\end{figure}

We have generated some plots for the scalar field $\phi $ given by Eq. (\ref%
{pp}) and the potential $V$ $\ $given by Eq.~(\ref{5}) with the help of Eq.~(%
\ref{kk}) for some numerical values of the constants $c_{1},c_{2},m,a_{1}$
(see Fig. (\ref{Fig1})).

Therefore, for the numerical values corresponding to the red and black lines
$V<0$,while for the blue line $V$ starts in a negative region, but goes to a
positive one. The parameters values for the magenta line give $V>0$. We have
chosen two numerical values for the exponent of the scale factor, $%
a_{1}=1/2, $ corresponding to a positive deceleration parameter, $q>0$, and
the value $a_{1}=3/2,$ indicating that the deceleration parameter is
negative, $q<0,$ that is, this choice represents an accelerating solution.

\subsection{Solution 2}

Again, by imposing the symmetry $\left[ c,\phi \right] $ where $c\in \mathbb{%
R},$ and using Eqs. (\ref{sys1})-(\ref{sys4}), we obtain the coupling
function $F\left( \phi \right) $, the Hubble parameter $H\left( t\right) $
and the scale factor $a(t)$ given by%
\begin{equation}
F\left( \phi \right) =m\phi ^{2},\qquad H\left( t\right) =a_{1},\qquad
a(t)=e^{a_{1}t}.  \label{ww}
\end{equation}%
It is possible to obtain a more general expression for $F$ from Eqs (\ref%
{sys1})-(\ref{sys4}) as
\begin{equation}
F=m\phi ^{2}-\frac{2e^{c_{1}c_{2}/2}}{c_{1}(c_{1}-4)}\phi ^{2-c_{1}/2}+c_{3}.
\end{equation}%
However, this expression for $F$ does not verify the field equations, and
this reason we have chosen only the particular solution $F=m\phi ^{2}.$
Then, by substituting these results into Eq.(\ref{eq2}), thus we get the
scalar field
\begin{eqnarray}
\phi &=&\phi _{0}\left\{ \frac{1}{ma_{1}}\left[ c_{1}\left( 8m-1\right)
e^{a_{1}t}+c_{2}\left( m-1\right) \right] \right\} ^{\frac{4m}{8m-1}},
\notag \\
&&c_{1},c_{2}\in \mathbb{R}^{-},\text{ \ }m<\frac{1}{8},\phi _{0}=4^{\frac{4m%
}{8m-1}},  \label{ff}
\end{eqnarray}%
while the invariant solution induced by the symmetry $\left[ c,\phi \right] $
yields the relation%
\begin{equation}
\phi =\phi _{0}e^{ct},  \label{xx}
\end{equation}%
immediately by inserting Eqs. (\ref{ww},\ref{xx}) into Eq. (\ref{eq2}) thus
we obtain the constant $c$ given by $c=\frac{4ma_{1}}{8m-1}$.

For simplicity, assume that $c_{2}=0$, then from Eq. (\ref{5}), we obtain
the potential $V$ as
\begin{equation}
V\sim e^{\frac{8ma_{1}}{8m-1}t},  \label{qw}
\end{equation}%
again the proportionality constant obtained from Eq. (\ref{qw}) is
complicated thus we shall not present its explicit form here.

\begin{figure}[h!]
\begin{center}
\includegraphics[height=2in,width=3in]{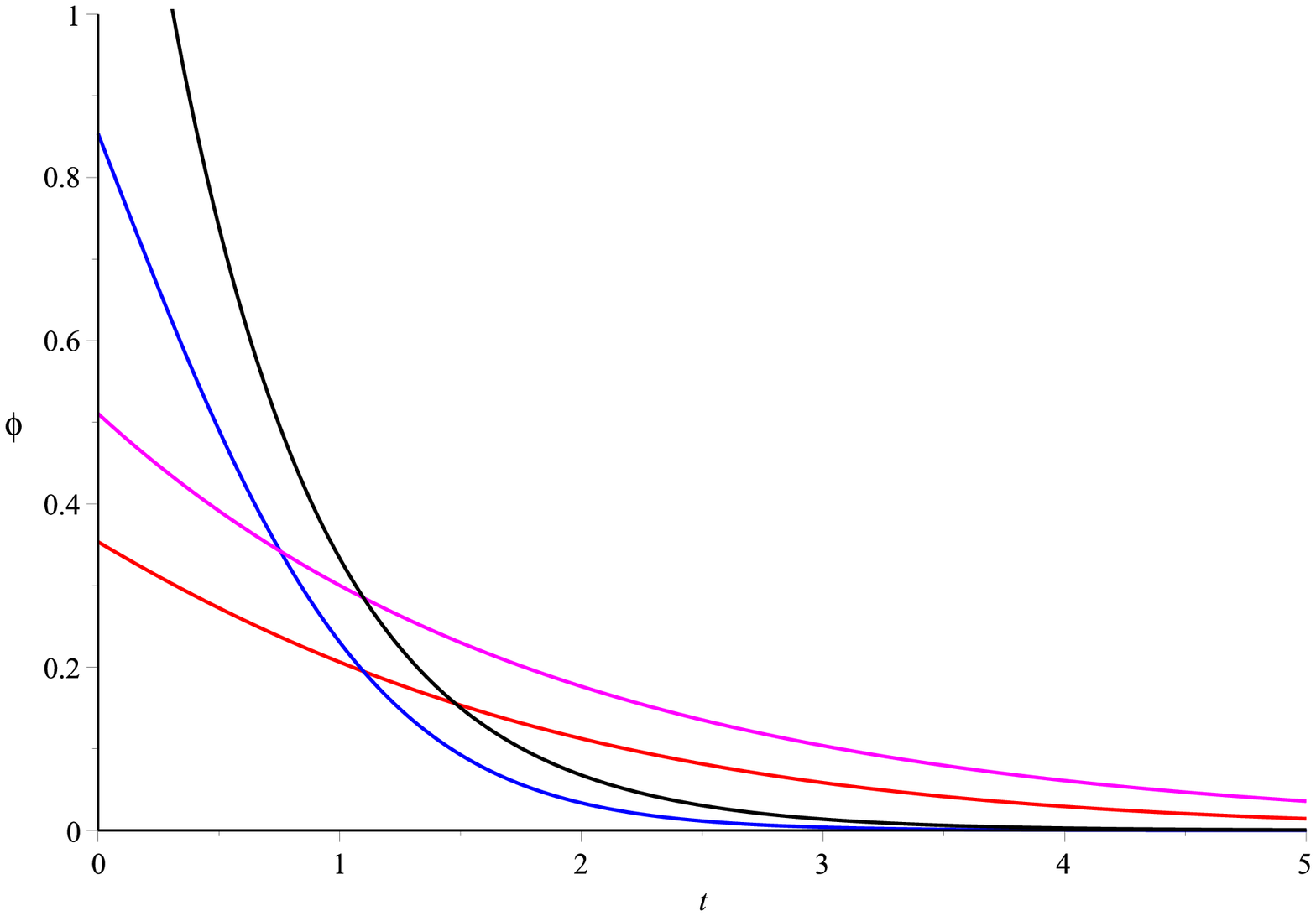} %
\includegraphics[height=2in,width=3in]{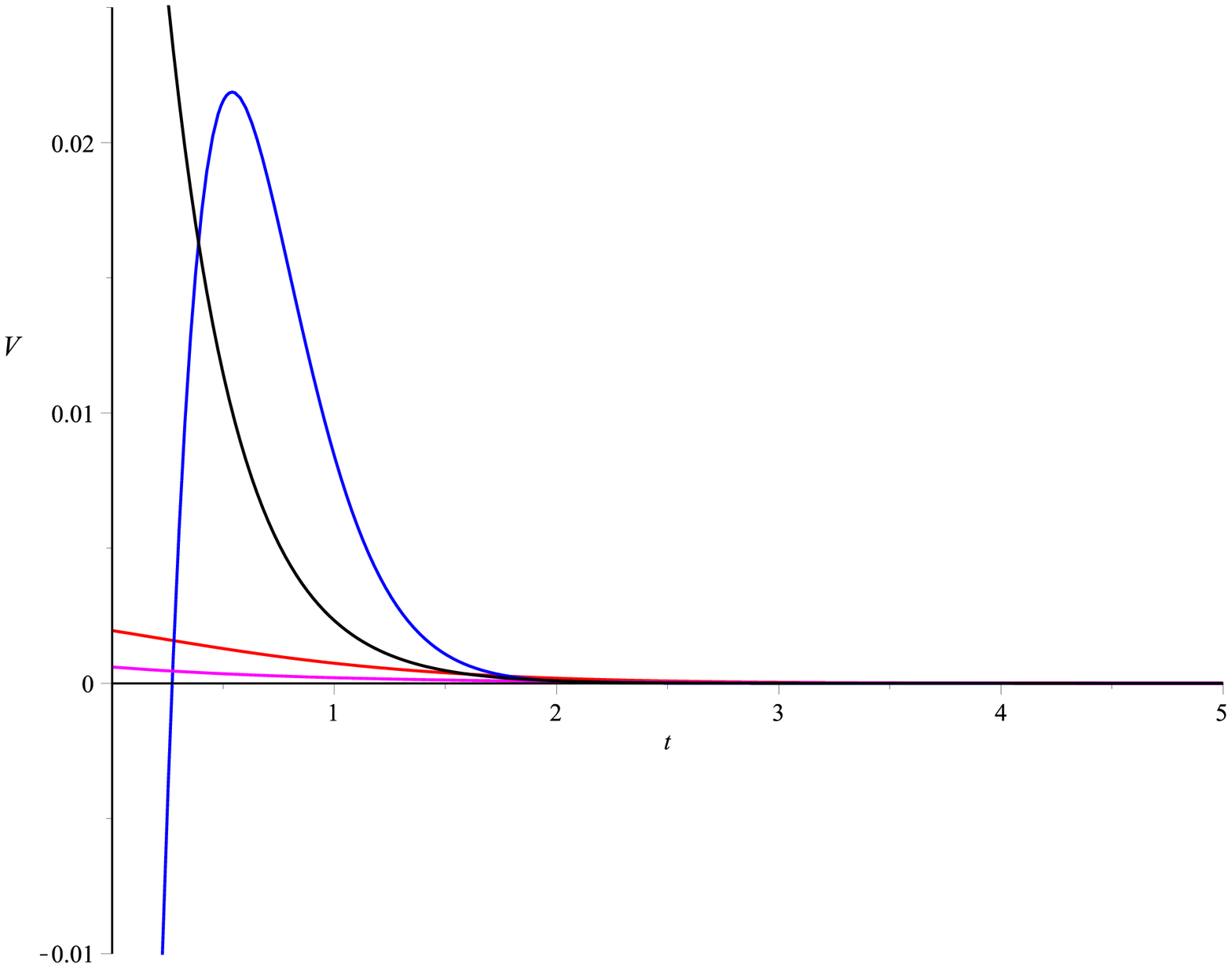}
\end{center}
\caption{Plots for the quantities $\protect\phi $ and $V$. Red line: $%
c_{1}=-1,c_{2}=-1,m=3/32,a_{1}=1/2$. Blue line: $%
c_{1}=-1,c_{2}=-1,m=3/32,a_{1}=3/2$. Magenta line: $%
c_{1}=-1,c_{2}=0,m=0.085,a_{1}=1/2$, and black line: $%
c_{1}=-1,c_{2}=0,m=0.085,a_{1}=3/2$ }
\label{Fig2}
\end{figure}

We have generated some plots for the scalar field $\phi $ given by Eq. (\ref%
{ff}) and the potential $V$ $\ $given by Eq.~(\ref{5}) with the help of Eq.~(%
\ref{ww}) for some numerical values of the constants $c_{1},c_{2},m$ and $%
a_{1}$ (see Fig. (\ref{Fig2}))

Therefore, for the parameters corresponding to the blue line $V$ starts in a
negative region, but goes into the positive one. Red, magenta and black
lines give $V>0.$ As before, we have chosen two numerical values for the
exponent of the scale factor, $a_{1}=1/2$ and $a_{1}=3/2,$ but the
deceleration parameter is always negative, $q=-1<0,$ that is, we have
obtained an accelerating solution.

In the next Section, we shall present the analytical solutions of the
differential Eq. (\ref{6}) by using Noether symmetries, subsequently leading
to some solutions of the gravitational field equation describing the
dynamics of the FRW universe in the presence of the scalar field in the
framework of the STT.

\section{Noether symmetries}

\label{sect5}

In order to obtain the solutions of the gravitational field equation through
the Noether symmetries, we first consider that the Lagrangian $\mathcal{L}$
is defined as $\mathcal{L}=\mathcal{L}\left( a,\dot{a},\phi ,\dot{\phi}%
\right) ,$ thus the configuration space $Q$ takes the form $Q=\left( a,\phi
\right) $. Note that the vector field $X$ is defined as
\begin{equation}
X=\xi \left( t,a,\phi \right) \partial _{t}+\eta \left( t,a,\phi \right)
\partial _{a}+\beta \left( t,a,\phi \right) \partial _{\phi },  \label{X0}
\end{equation}%
where we have denoted the arbitrary functions $\xi $,$\eta $ and $\beta $ as
$\xi =\xi \left( t,a,\phi \right) ,$ $\eta =\eta \left( t,a,\phi \right) ,$ $%
\beta =\beta \left( t,a,\phi \right) $ respectively. Now the first
prolongation formula for the vector field $X$ is given by%
\begin{equation}
X^{\left[ 1\right] }=X+\dot{\eta}\partial _{\dot{a}}+\dot{\beta}\partial _{%
\dot{\phi}},  \label{X1}
\end{equation}%
where we have introduced the following notations
\begin{equation}
\dot{\eta}=D_{t}\eta -\dot{a}D_{t}\xi ,\dot{\beta}=D_{t}\beta -\dot{\phi}%
D_{t}\xi ,D_{t}=\partial _{t}+\dot{a}\partial _{a}+\dot{\phi}\partial _{\phi
}.
\end{equation}%
It is worth to note that the vector field $X$ can generate a Noether
symmetry if the Lagrangian $\mathcal{L}$ and the function $A$ defined as $%
A=A\left( t,a,\phi \right) $ satisfy the relation (see for insatnce \cite%
{Bluman},\cite{olver},\cite{Ibragimov} and \cite{Cantwell})
\begin{equation}
X^{\left[ 1\right] }\mathcal{L}+D_{t}\left( \xi \right) \mathcal{L}=D_{t}A.
\label{EQ}
\end{equation}%
For the specific case $A=0$, the vector field $X$ generates the classical
Noether symmetry. From the mathematical point of view, the conserved
quantity $I$ can be obtained from%
\begin{equation}
I=\sum_{i}\left( \alpha _{i}-\xi \dot{q}_{i}\right) \frac{\partial \mathcal{L%
}}{\partial \dot{q}_{i}}+\xi \mathcal{L}-A,  \label{Inv}
\end{equation}%
where $\alpha _{1}=\xi ,$ $\alpha _{2}=\eta $ and $\alpha _{3}=\beta .$ In
view of Eqs. (\ref{2}) and (\ref{EQ})$,$ we get a system of the partial
differential equations
\begin{widetext}
\begin{align}
\partial _{a}\xi & =0=\partial _{\phi }\xi , \label{n}  \\
F\eta +F^{\prime }a\beta +2Fa\partial _{a}\eta +F^{^{\prime }}a^{2}\partial
_{a}\beta -Fa\partial _{t}\xi & =0,   \\
\frac{3}{2}a^{2}\eta +6F^{^{\prime }}a^{2}\partial _{\phi }\eta
+a^{3}\partial _{\phi }\beta -\frac{1}{2}a^{3}\partial _{t}\xi & =0,
\\
12F^{^{\prime }}a\eta +6F^{\prime \prime }a^{2}\beta +12Fa\partial _{\phi
}\eta +6F^{^{\prime }}a^{2}\partial _{a}\eta +6F^{^{\prime }}a^{2}\left(
\partial _{\phi }\beta -\partial _{t}\xi \right) +a^{3}\partial _{a}\beta &
=0,   \\
2Fa\partial _{t}\eta +F^{^{\prime }}a^{2}\partial _{t}\beta & =\partial
_{a}A,  \\
6F^{^{\prime }}a^{2}\partial _{t}\eta +a^{3}\partial _{t}\beta & =\partial
_{\phi }A,  \\
-3a^{2}V\eta -a^{3}V^{\prime }\beta -a^{3}V\partial _{t}\xi & =\partial
_{t}A.  \label{aa}
\end{align}
\end{widetext}

\subsection{Solution 1 with $A=0$}

With the help of Eqs.~(\ref{n})-(\ref{aa}), we obtain the following results%
\begin{eqnarray}
F &=&m\phi ^{2},\quad \eta =-\frac{2}{3}ac_{2},\quad \beta =c_{2}\phi ,\quad
\notag \\
&&V=V_{0}\phi ^{2},\quad \xi =0,\qquad m,c_{2},V_{0}\in \mathbb{R}.
\label{ab}
\end{eqnarray}%
The conserved quantity $I$ is given by Eq. (\ref{Inv}), where in case it
reads
\begin{equation}
I=\eta \frac{\partial \mathcal{L}}{\partial \dot{a}}+\beta \frac{\partial
\mathcal{L}}{\partial \dot{\phi}}.  \label{as}
\end{equation}%
By substituting Eqs. (\ref{2},\ref{ab}) into Eq. (\ref{as}), the latter
yields the following differential equation%
\begin{equation}
I=c_{2}\left[ 4m\phi ^{2}a^{2}\dot{a}+\left( 1-8m\right) a^{3}\phi \dot{\phi}%
\right] .  \label{s1}
\end{equation}%
In order to find the scalar field $\phi $ explicitly, thus we rewrite Eq. (%
\ref{s1}) as a Bernoulli differential equation for $\phi ^{2}$ in the form

\begin{equation}
\frac{d}{da}\phi ^{2}+\frac{8m}{8m-1}\frac{1}{a}\phi ^{2}=\frac{2I}{%
c_{2}\left( 1-8m\right) a^{3}\frac{da}{dt}},  \label{s2}
\end{equation}%
Eq. (\ref{s2}) can be integrated to give the scalar field
\begin{equation}
\phi _{\pm }\left( a,t\right) =\pm a^{\frac{4m}{8m-1}}\sqrt{C_{1\pm }+\frac{2%
}{c_{2}\left( 1-8m\right) }\int Ia^{\frac{3-32m}{8m-1}}dt},  \label{s3}
\end{equation}%
where $C_{1\pm }$ are the arbitrary constants of integration. Consider that
the conserved quantity $I$ is a constant and $m$ is $3/32$, then from Eq. (%
\ref{s3}), we get the result
\begin{equation}
\phi _{\pm }\left( a,t\right) =\pm a^{-\frac{3}{2}}\sqrt{C_{1\pm }+\frac{8I}{%
c_{2}}t}.
\end{equation}%
Furthermore, if the conserved quantity $I$ vanishes, then from Eq. (\ref{s3}%
) we obtain the relation

\begin{equation}
\phi _{\pm }\left( a\right) =\pm \sqrt{C_{1\pm }}a^{\frac{4m}{8m-1}}.
\end{equation}

The invariant solution is given by $\frac{da}{a}=-\frac{2d\phi }{3\phi }$,
thus we get: $\phi =a^{-3/2}$. By substituting this result together with $%
F=m\phi ^{2}$ and $V=V_{0}\phi ^{2}$, for example, from Eq.(\ref{eq2}), we
obtain the relation $\phi =e^{-Kt}$ where $K$ is a constant that depends on $%
m.$Therefore this invariant solution is the same as the invariant one
obtained in the second solution through the Lie group method. Note that the
Noether symmetries generate an algebra which is a subalgebra of the Lie
algebra.

Next in order to obtain the general solutions of the gravitational field
equations, we introduce two arbitrary functions $w\left( a,\phi \right) $
and $z\left( a,\phi \right) $ induced by the vector field $X$ defined as%
\begin{equation}
w\left( a,\phi \right) =a^{3/2}\phi ,\quad z\left( a,\phi \right) =-\frac{3}{%
2}\ln a+a^{3/2}\phi ,  \label{z0}
\end{equation}%
respectively. For mathematical convenience, we rewrite Eqs. (\ref{z0}) in
the forms
\begin{equation}
a\left( w,z\right) =e^{\frac{2}{3}\left( w-z\right) },\qquad \phi \left(
w,z\right) =we^{z-w}.  \label{z2}
\end{equation}

\begin{figure}[h]
\begin{center}
\includegraphics[height=1.5in,width=2.25in]{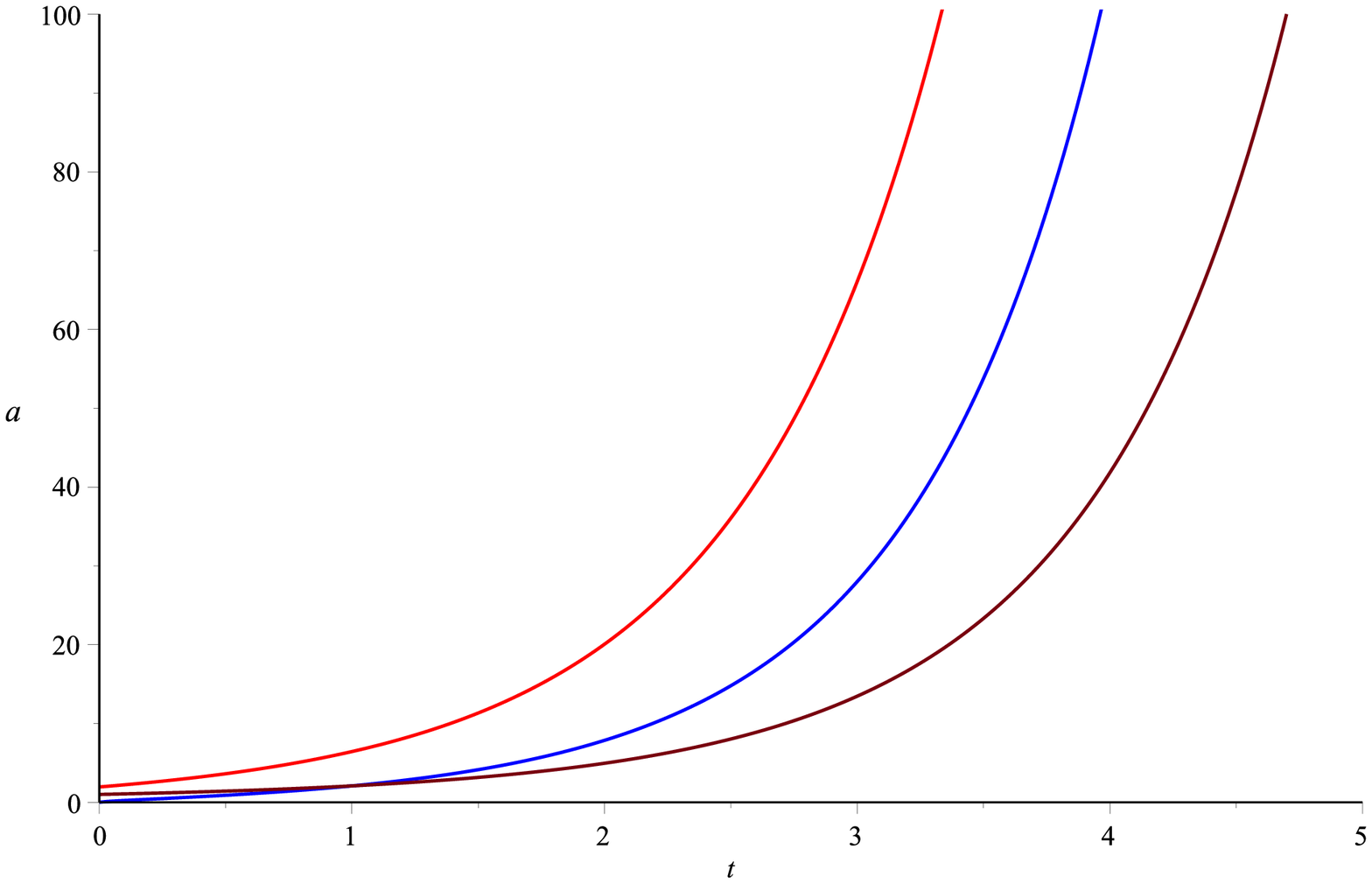} %
\includegraphics[height=1.5in,width=2.25in]{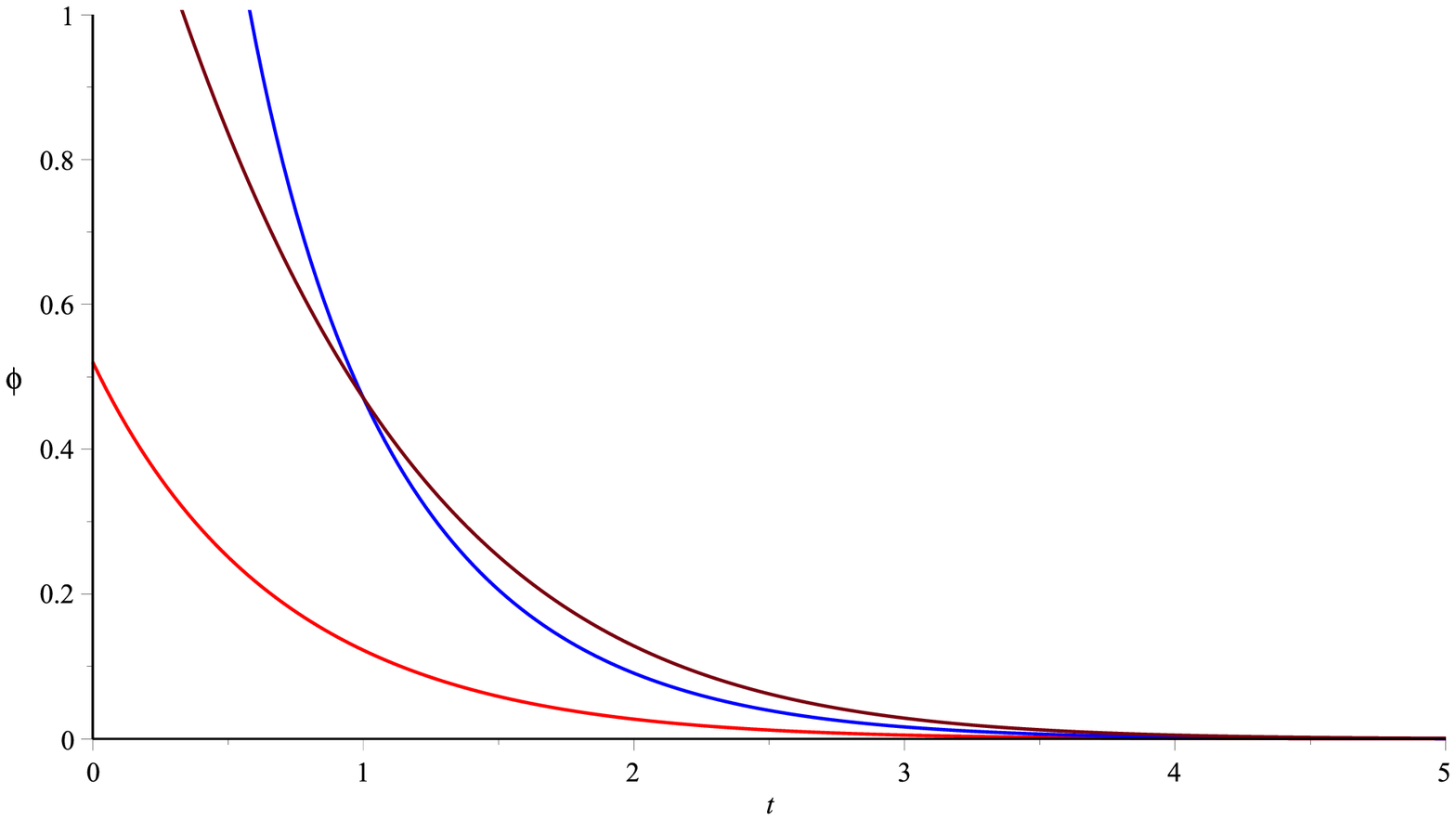} %
\includegraphics[height=1.5in,width=2.25in]{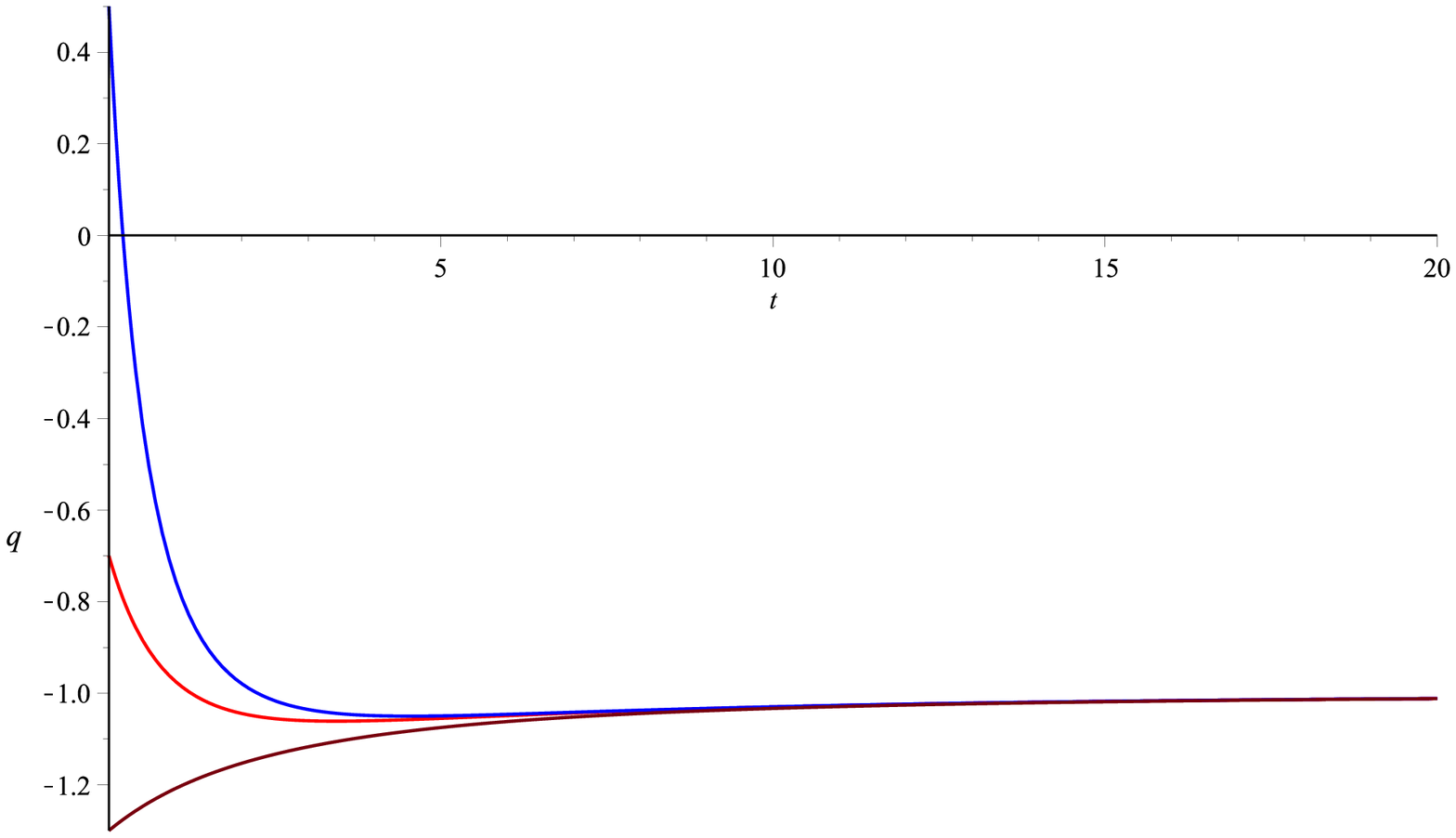}
\end{center}
\caption{Plots of the quantities $a,\protect\phi ,q$. With $V_{0}=1/40$. Red
line: $c=1,b=1,c_{1}=-1,c_{2}=-1$. Blue line: $c=1,b=0,c_{1}=-1,c_{2}=0$ and
black color: $c=0,b=1,c_{1}=-1,c_{2}=0$}
\label{Fig3}
\end{figure}

Now by inserting the relations $F=m\phi ^{2}$, $V=V_{0}\phi ^{2}$, $m=3/32$
and Eqs. (\ref{z0},\ref{z2}) into Eq. (\ref{2}), then the Lagrangian (\ref{2}%
) takes the simpler form
\begin{equation}
\mathcal{L}=\frac{\dot{w}w}{4}\left( \dot{z}-\dot{w}\right) +\frac{\dot{w}%
^{2}}{2}-V_{0}w^{2}.
\end{equation}%
Note that the variable $z$ here is cyclic. Next using the Euler-Lagrangian
equations, thus we obtain the arbitrary functions $\ w\left( t\right) $ and $%
z\left( t\right) $ given by
\begin{widetext}
\begin{equation}
w\left( t\right) =\sqrt{ct+b},\qquad z\left( t\right) =-\ln \left\vert
ct+b\right\vert +\left( ct+b\right) ^{1/2}-4V_{0}t^{2}+c_{1}t+c_{2}.
\end{equation}%
\end{widetext}. It is easy to obtain the solutions in the original variables%
\begin{equation}
a\left( t\right) =\left( ct+b\right) ^{2/3}e^{\frac{2}{3}\left(
4V_{0}t^{2}-c_{1}t-c_{2}\right) },\phi \left( t\right) =\frac{%
e^{-4V_{0}t^{2}+c_{1}t+c_{2}}}{\sqrt{ct+b}}.
\end{equation}%
We have plotted the quantities $a,\phi $ and the deceleration parameter $q$
defined as $q=\frac{d}{dt}\left( \frac{1}{H}\right) -1$, for different
values of the constants $c,b,c_{1},c_{2}$ and $V_{0}$. See Fig. (\ref{Fig3}).

As one can see from the Figures, all the plotted solutions have an
inflationary (accelerating) behavior.

\subsection{Solution 2 with $A=0$}

This solution has been already obtained one by Capozziello and the Ritis
\cite{1c}, with $A=0:$
\begin{equation}
F=m\phi ^{2},\eta =-c_{2}Ka^{n_{1}}\phi ^{n_{2}},\beta =c_{2}a^{b_{1}}\phi
^{b_{2}},V=V_{0}\phi ^{\lambda },
\end{equation}%
where%
\begin{align}
K& =\frac{\sqrt{3}\left( 2\sqrt{3m}+\sqrt{12m-1}\right) }{6\sqrt{m}},n_{1}=-%
\frac{1}{2}-\frac{\sqrt{3m}}{\sqrt{12m-1}},  \notag \\
n_{2}& =-1+\frac{\sqrt{3}\left( 8m-1\right) }{4\sqrt{12m^{2}-m}},b_{1}=-9-%
\frac{2\sqrt{3m}}{\sqrt{12m-1}},  \notag \\
b_{2}& =\frac{\sqrt{3}\left( 8m-1\right) }{4\sqrt{m}\sqrt{12m-1}},\lambda =3+%
\frac{\sqrt{36m-3}}{2\sqrt{m}}.  \notag
\end{align}%
In \cite{1c} the authors studied in particular the case $m=\frac{3}{32}.$ We
have only obtained another parametrization by using a Lie point symmetry
approach instead of the geometrical one followed from \cite{1c} (see also
\cite{13}).

\subsection{Solution 3 with $A\neq 0$}

A third set of solutions corresponds to the choices%
\begin{eqnarray}
F &=&m\phi ^{2},V=\lambda \phi ^{n},\xi =c_{1}t+c_{2},  \notag \\
\eta &=&\frac{\left( n+2\right) c_{1}}{3\left( n-2\right) }a,\beta =-\frac{%
2c_{1}}{n-2}\phi ,A=c_{3}.  \label{K}
\end{eqnarray}
Since $n\neq 2,$ it follows that
\begin{equation}
X=\left( c_{1}t+c_{2}\right) \partial _{t}+\frac{\left( n+2\right) c_{1}}{%
3\left( n-2\right) }a\partial _{a}-\frac{2c_{1}}{n-2}\phi \partial _{\phi },
\end{equation}%
and therefore we obtain%
\begin{equation}
X_{1}=t\partial _{t}+\frac{n+2}{3\left( n-2\right) }a\partial _{a}-\frac{2}{%
n-2}\phi \partial _{\phi },\qquad X_{2}=\partial _{t}.
\end{equation}%
As one can observe easily, $X_{1}$ induces the following invariant solution,
\begin{equation}
a=a_{0}t^{\frac{n+2}{3\left( n-2\right) }},\qquad \phi =\phi _{0}t^{-\frac{2%
}{n-2}},
\end{equation}%
which is the self-similar solution already obtained by using the Lie group
method (Solution 1), since it verifies the relations: $F^{-1}V=t^{-2},$ and $%
F^{-1}\dot{\phi}^{2}=t^{-2}.$ Now using Eq. (\ref{Inv}) and $c_{1}=1$, $%
c_{2}=0$, thus it is clear to write down the conserved quantity

\begin{equation}
I_{1}=\left[ \frac{n+2}{3\left( n-2\right) }a-\dot{a}t\right] \frac{\partial
\mathcal{L}}{\partial \dot{a}}+\left( \frac{2}{2-n}-\dot{\phi}t\right) \frac{%
\partial \mathcal{L}}{\partial \dot{\phi}}+t\mathcal{L}-A,  \label{c}
\end{equation}%
for calculational convenience, we rewrite Eq. (\ref{c}) in the form

\begin{eqnarray}
I_{1}&=&\frac{1}{n-2}\left( \frac{n+2}{3}a\frac{\partial \mathcal{L}}{%
\partial \dot{a}}-2\phi \frac{\partial \mathcal{L}}{\partial \dot{\phi}}%
\right) +  \notag \\
&&t\left( \mathcal{L}-\dot{a}\frac{\partial \mathcal{L}}{\partial \dot{a}}-%
\dot{\phi}\frac{\partial \mathcal{L}}{\partial \dot{\phi}}\right) -A.
\label{ar}
\end{eqnarray}
With the help of Eqs. (\ref{2}) and (\ref{K}), Eq. (\ref{ar}) takes the form

\begin{eqnarray}
I_{1} &=&\frac{2ma^{2}\phi }{n-2}\left[ 2\left( n-4\right) \phi \dot{a}%
+\left( 2n+3\right) a\dot{\phi}\right] +  \notag \\
&&t\left[ -6ma\dot{a}\phi \left( \phi \dot{a}+2a\dot{\phi}\right) -\frac{1}{2%
}a^{3}\dot{\phi}^{2}-a^{3}\lambda \phi ^{n}\right] -A.
\end{eqnarray}%
Similarly, another conserved quantity $I_{2}$ takes the form

\begin{equation}
I_{2}=-\dot{a}\frac{\partial \mathcal{L}}{\partial \dot{a}}-\dot{\phi}\frac{%
\partial \mathcal{L}}{\partial \dot{\phi}}+\mathcal{L}-A.
\end{equation}%
In view of Eq. (\ref{2}), thus we obtain the relation

\begin{equation}
I_{2}=-6ma\dot{a}\phi \left( \phi \dot{a}+2a\dot{\phi}\right) -\frac{1}{2}%
a^{3}\dot{\phi}^{2}-a^{3}\lambda \phi ^{n}-A,
\end{equation}%
explicitly. Note that from Eq. (\ref{5}), we get
\begin{equation}
-6ma\dot{a}\phi \left( \phi \dot{a}+2a\dot{\phi}\right) -\frac{1}{2}a^{3}%
\dot{\phi}^{2}-a^{3}\lambda \phi ^{n}=0,
\end{equation}%
and therefore the relation $I_{2}=-A$ holds$.$ Thus we get the result
\begin{equation}
\tilde{I}_{1}=\frac{2ma^{2}\phi }{n-2}\left[ 2\left( n-4\right) \phi \dot{a}%
+\left( 2n+3\right) a\dot{\phi}\right] ,  \label{m1}
\end{equation}%
where we have denoted $\tilde{I}_{1}=I_{1}+A$. In order to solve Eq. (\ref%
{m1}) for the scalar field $\phi $, thus we rewrite Eq. (\ref{m1}) as a
Bernoulli differential equation for $\phi ^{2}$ in the form

\begin{equation}
\frac{d}{da}\phi ^{2}+\frac{4\left( n-4\right) }{2n+3}\frac{1}{a}\phi ^{2}=%
\frac{n-2}{\left( 2n+3\right) m}\frac{\tilde{I}_{1}}{a^{3}\frac{da}{dt}},
\label{m2}
\end{equation}%
Eq. (\ref{m2}) can be integrated to give the scalar field
\begin{equation}
\phi _{\pm }\left( a,t\right) =\pm a^{\frac{2\left( 4-n\right) }{2n+3}}\sqrt{%
C_{2\pm }+\frac{n-2}{\left( 2n+3\right) m}\int \tilde{I}_{1}a^{-\frac{2n+25}{%
2n+3}}dt},  \label{m3}
\end{equation}%
where $C_{2\pm }$ are the arbitrary constants of integration. Consider that
the quantity $\tilde{I}_{1}$ vanishes, that is $I_{1}=-A=-c_{3}$ then from
Eq. (\ref{m3}), we get the relation

\begin{equation}
\phi _{\pm }\left( a\right) =\pm \sqrt{C_{2\pm }}a^{\frac{2\left( 4-n\right)
}{2n+3}}.  \label{m4}
\end{equation}%
For the special case $n=4$, then the scalar field $\phi $ becomes an
arbitrary constant, thus from Eq. (\ref{5}), we obtain the scale factor as

\begin{equation}
a(t)=e^{\left[ \frac{\sqrt{-6m\lambda c_{1}}}{6m}\left( t-c_{2}\right) %
\right] }.
\end{equation}%
In order to have the realistic cosmological model, therefore we have chosen
the relations $c_{1}<0$, $\lambda >0$ and $m>0$.

\section{Conclusions}

\label{sect6}

Scalar fields play an important role in the explanation of the observational
aspects of present day cosmology. In particular, scalar fields are
responsible for the early inflationary expansion of the Universe, and they
also represent a powerful candidate for the dark energy determining the
recent accelerated expansion of the Universe. In the present work, by using
the standard procedures, Lie group approach and Noether symmetry techniques,
we have obtained several analytical solutions of the gravitational field
equations describing the time evolutions of a flat Friedman-Robertson-Walker
Universe in the presence of scalar fields.

With regard to the Noether symmetry method, in the present paper we have employed an approach, that
allows us to obtain more symmetries thanks to the function $A$ (a boundary
term introduced into the action to keep the action invariant by $X$). An alternative method is
the geometrical one (see \cite{Cappo1996} for
an excellent review of this method). To obtain exact cosmological solutions
in generalized gravity theories, which can be, for example,  scalar-tensor theories with a
non-minimally (or minimally) coupled scalar field, or higher-order theories (that is
theories whose Lagrangian is not simply linear in Ricci scalar), one can use a method
specifically developed to this purpose,  which is called the Noether Symmetry Approach \cite{Cappo1996}. This method can
be described as follows. If a (point-like) Lagrangian is given in terms of some variables (i.e., in the cosmological case, in terms of
the scale factor of the Universe $a$ and a scalar field $\phi$), one can  search for a Noether
symmetry of such a Lagrangian, which contains an a priori unspecified potential $V$, or,
in the non-minimal coupling case, an unspecified coupling $F(\phi)$, and a potential $V(\phi)$, respectively.
Hence in this approach one relates the possible existence of the Noether symmetry
(that is, the existence of a constant of motion) with the selection of specific forms of
$V(\phi)$, or of $F(\phi)$ and $V(\phi)$, respectively. In this way, the problem of solving the model is connected
with the specification of $V(\phi)$ (or $F(\phi)$ and $V(\phi)$). The existence of symmetries selects couplings
and potentials of physical interest as $V(\phi)=\lambda \phi ^4$ or $F(\phi) =k_0\phi ^2$ \cite{Cappo1996}, respectively, with $\lambda $ and $k_0$ constants.

Hence by using Noether symmetry approaches a large number of cosmological models
of physical interest can be found, and their properties can be tested with the astrophysical
observations.

The solutions we have found show
a large variety of cosmological behaviors, ranging from non-accelerating
(decelerating) solutions to accelerating ones. The scalar field potential
can be also obtained, and it is basically determined by symmetry and/or
invariance considerations. Our solutions can also be
interpreted as the background on which to compare the observational data from
large-scale structure formation, from the Cosmic Microwave Background Radiation anisotropies \cite{6}, or on which to formulate a theory of cosmological perturbations in modified gravity models. For example, our solutions show that it is possible to control the cosmological
background by a free parameter, so that it is possible to select interesting scales
useful for large-cosmological-structure formation. Generally, the potentials have either a power law
type dependence on the scalar field, like in the case of the first analytic
solution, or their time dependence can be obtained explicitly. In this
latter case we also have a large variety of behaviors, with both power law
and exponential time evolution of $V$. Hence the results of the present
study indicate that simple exact cosmological models can be obtained in the
framework of scalar field cosmologies by using the powerful methods of Lie
group analysis, and Noether symmetry approach. The cosmological implications
of the present results will be investigated in detail in a future
publication.

\section*{Acknowledgments} We would like to thank the anonymous reviewer for comments and suggestions that helped us to improve our manuscript.

\end{document}